\documentclass{cjaa}                   

\usepackage{graphicx}                   
\input{epsf.sty}                        

\begin{document}

   \title{Distribution of $^{56}$Ni Yields of Type Ia Supernovae
   and Implications for Progenitors}

   \setcounter{page}{1}          

   \author{Bo Wang
      \inst{1,2}\mailto{}
   \and Xiang-Cun Meng
      \inst{1,2}
    \and Xiao-Feng Wang
      \inst{3,4}
   \and Zhan-Wen Han
      \inst{1}
      }
   \offprints{B. Wang}                   
   \baselineskip=4.642mm   

   \institute{National Astronomical Observatories/Yunnan
Observatory, Chinese Academy of Sciences, Kunming 650011\\
             \email{wangbo@ynao.ac.cn}
        \and
             Graduate School of Chinese Academy of Sciences, Beijing 100049\\
        \and
        Tsinghua Center for Astrophysics (THCA) and Department of
        Physics, Tsinghua University, Beijing 100084\\
        \and
        Astronomy Department, University of California at
        Berkeley, CA 94720, USA}

   \date{22~~June~~2007}

   \abstract{
   The amount of $^{56}$Ni produced in type Ia supernova (SN Ia) explosion
   is probably the most important physical parameter underlying the observed
   correlation of SN Ia luminosities with their light curves. Based on an
   empirical relation between the $^{56}$Ni mass and the light curve parameter
   $\triangle m_{15}$, we obtained rough estimates of the $^{56}$Ni mass for a
   large sample of nearby SNe Ia with the aim of exploring the diversity in SN Ia.
   We found that the derived $^{56}$Ni masses for different SNe Ia could vary by
   a factor of ten (e.g., $M_{\rm Ni}=0.1 - 1.3$\,$M_{\odot}$), which cannot
   be explained in terms of the standard Chandrasekhar-mass model (with a $^{56}$Ni
   mass production of 0.4 -- 0.8\,$M_{\odot}$). Different explosion and/or progenitor
   models are clearly required for various SNe Ia, in particular, for those extremely
   nickel-poor and nickel-rich producers. The nickel-rich (with $M_{\rm Ni}$ $>$
   0.8\,$M_{\odot}$) SNe Ia are very luminous and may have massive progenitors
   exceeding the Chandrasekhar-mass limit since extra progenitor fuel is required
   to produce more $^{56}$Ni to power the light curve. This is also consistent
   with the finding that the intrinsically bright SNe Ia prefer to occur in
   stellar environments of young and massive stars. For example, 75$\%$ SNe Ia
   in spirals have $\Delta m_{15} < 1.2$ while this ratio is only 18$\%$ in E/S0
   galaxies. On the other hand, the nickel-poor SNe Ia (with $M_{\rm Ni}$ $<$
   0.2\,$M_{\odot}$) may invoke the sub-Chandrasekhar model, as most of them
   were found in early-type E/S0 galaxies dominated by the older and low-mass
   stellar populations. This indicates that SNe Ia in spiral and E/S0 galaxies
   have progenitors of different properties.
   \keywords{stars: evolution --- supernovae : general --- white dwarfs}
   }

   \authorrunning{B. Wang et al.}            
   \titlerunning{Distribution of $^{56}$Ni Yields}  

   \maketitle

%
%

\section{Introduction}           
\label{sect:intro}
It is well known that Type Ia Supernovae (SNe Ia) are excellent
cosmological distance indicators due to their high luminosity and
remarkable uniformity. The results of the High-Z Supernova Search
Team (Riess et al. 1998) and the Supernova Cosmology Project
(Perlmutter et al. 1999) astonishingly showed that the expansion of
the universe is accelerating. Although SNe Ia are very important
objects in modern cosmology, several key issues related to the
nature of their progenitor systems (Hachisu et al. 1996, 1999; Li
$\&$ van den Heuvel 1997; Han $\&$ Podsiadlowski 2004; Meng et al.
2006) and the physics of the explosion mechanisms (Hillebrandt $\&$
Niemeyer 2000; R\"{o}pke $\&$ Hillebrandt 2005a) are still not
understood.

It is widely accepted that SNe Ia are thermonuclear explosions of
carbon-oxygen (C-O) white dwarfs (WDs) accreting matter from their
companions (for a review see Nomoto et al. 1997). The energy
released from the burning till nuclear statistical equilibrium (NSE)
completely destroys the C-O WD. The optical/IR light curves are
powered by the radioactive decay of $^{56}$Ni $\rightarrow$
$^{56}$Co $\rightarrow$ $^{56}$Fe (Colgate $\&$ Mckee 1969).

Over the last decades, it has been found that there exists
spectroscopic diversity among SNe Ia. In an attempt to quantify the
rate of spectroscopically peculiar SNe Ia in the existing observed
sample, Branch et al. (1993) compiled a set of 84 SNe Ia and found
that about $83\% - 89\%$ of the sample are normal (or sometimes
``Branch normals''). According to Li et al. (2001), however, only
$64\%\pm 12\%$ of observed SNe Ia are normal in a volume-limited
search. The total rate of peculiar SNe Ia could be as high as
$36\%\pm 9\%$. The rates are $16\%\pm 7\%$ and $20\%\pm 7\%$ for SN
1991bg-like events and SN 1991T-like objects, respectively.
Moreover, some of the peculiar SNe Ia apparently deviate from the
relation between the light-curve shape parameter $\Delta m_{15}$ and
the luminosity (e.g., Wang et al. 2006), which probably suggests
that SN Ia luminosity is not a single parameter of the light curve
shape. Note that Wang et al. (2005) presented a single post-maximum
color parameter $\triangle C_{12}$ ($B - V$ color $\sim$\,12 days
after the $B$-band light maximum), which empirically describes
almost the full range of the observed SN Ia luminosities and gives
tighter correlations with SN Ia luminosities, but the underpinning
physics is still not understood. Stritzinger et al. (2006a) recently
noticed that no explosion model has been proposed yet to
successfully account for the factor of ten spanned by the range in
the observed bolometric luminosity.

In this paper, we aim to understand the origin of the progenitors
and explosion models with a large well-observed sample of nearby SNe
Ia. According to a simple empirical relation between the $^{56}$Ni
mass ($M_{\rm Ni}$) and the decline rate parameter $\triangle
m_{15}$, we can estimate the mass of radioactive material $^{56}$Ni
synthesized in SN Ia explosions. By studying the distribution of
such a parameter for SNe Ia, we hope to obtain some clues to the SN
Ia diversities. Moreover, we also investigate the correlation of SNe
Ia distribution with their host galaxies, which might also set
constraints on the SN Ia progenitors.

The paper is organized as follows. In Section 2, we provide a brief
description of the observational data examined in this study.
Section 3 shows the distribution of the decline rate parameter
$\triangle m_{15}$, the relative radial distance $r_{\rm SN}/R_{25}$
and the derived $^{56}$Ni mass. In particular, a simple empirical
method is introduced and used to estimate the $^{56}$Ni mass for a
large number of well-observed SNe Ia. A discussion and conclusions
are then given in Section 4.


\section{DATA}
\label{sect:data}
In this paper, we consider a sample of 111 well-observed nearby SNe
Ia with $Z<0.1$, of these 109 were compiled by Wang et al. (2006)
(see tables 1 and 2 in their work), and SNe 2002cs and 2002dj are
from unpublished KAIT SN database. The sample (excepting SNe 1937C,
1972E and 1974G) includes only SNe with CCD measurements and those
observed not later than 8 days after the \emph{B}-band light maximum
and with more than five photometric points. A large fraction of
these observations were obtained by the earlier Calan/Tololo SN
survey, the CfA I/II SN monitoring campaign, and the Las
Campanas/CTIO observing campaign.

The sample contains two important parameters (i.e., $\triangle
m_{15}$ and $r_{\rm SN}/R_{25}$). The decline rate parameter
$\triangle m_{15}$ was defined by Phillips (1993) as the amount in
magnitudes that the \emph{B}-band light curve decays in the first 15
days after maximum light, and the decline rate was corrected for a
small reddening effect (Phillips et al. 1999). $r_{\rm SN}/R_{25}$
is the de-projected galactocentric distance of the SNe in their
respective host galaxies in units of the galaxy radius $R_{25}$,
where $r_{\rm SN}$ is the de-projected radial distance of the SN Ia
from the galactic center and $R_{25}$ is the apparent semimajor axes
of the SN host galaxy isophote having
$\mu_{B}=25.0\rm\,mag\,arcsec^{-2}$. The 111 well-observed sample
contains 90 spectroscopically normal SNe Ia, 13 SN 1991T-like
objects, six SN 1991bg-like events and two truly peculiar SNe Ia,
SNe 2001ay and 2002cx. The 111 sample can be divided into four
groups by the morphological type of their host galaxies: 77 are in
spiral galaxies, 28 in E/S0 galaxies, two (SNe 1972E and 2000fa) in
irregular galaxies, and four (SNe 1992bl, 1999aw, 1999ej and 2002cx)
in galaxies of uncertain or intermediate types. Again, the sample of
111 contains 98 SNe Ia with relative radial distances while the
remainder are uncertain. Out of the 98 SNe Ia with known distances,
26 are from E/S0 galaxies, 70 from spiral galaxies, one (SN 1992bl)
from a S0/a galaxy and one (SN 2000fa) from an irregular galaxy.

\section{ANALYSIS AND RESULTS}
\label{sect:analysis}
In this section we investigate the distributions of $\triangle
m_{15}$ and the derived $^{56}$Ni mass versus $r_{\rm SN}/R_{25}$.
In particular, we provide a short overview of the method to
estimate the $^{56}$Ni masses for a large number of well-observed
SNe Ia.

\subsection{Distribution of $\triangle m_{15}$}
Using optical light curves of SNe Ia, Phillips (1993), Hamuy et al.
(1996a) and Phillips et al. (1999) established a relationship
between the absolute magnitude at \emph{B}-band light maximum and
the initial decline rate of the light curve. The characteristic
parameter is $\triangle m_{15}$. When $\triangle m_{15}$ is properly
corrected for, the SNe Ia prove to be a high precision distance
indicator, yielding relative distances with uncertainties $\sim$
7$\%$ -- 10$\%$ (Hamuy et al. 1996b). Kasen $\&$ Woosley (2007)
confirmed that the brighter SNe Ia have broader $B$-band light
curves with smaller decline rates around optical maxima than the
dimmer SNe Ia. Moreover, the determination of the decline rate
parameter $\triangle m_{15}$ is available for a large number of
objects. By analyzing $\triangle m_{15}$, it is possible for us to
peer into some intrinsic properties of SNe Ia, because it is
basically independent of distance and reddening.

Hamuy et al. (1995, 1996a) first noticed that brighter (also smaller
$\triangle m_{15}$) SNe Ia are preferentially located in late-type
galaxies for the nearby sample of SN Ia (see also Gallagher et al.
2005). In Figure 1, we show the distribution of $\triangle m_{15}$
for spiral and E/S0 galaxies, where the mean error for $\triangle
m_{15}$ is 0.07. It is obvious that the distribution of $\triangle
m_{15}$ is different in spiral and E/S0 galaxies. A K-S test shows
that the probability of these two distributions coming from the same
population is as low as $P=3.3\times10^{-7}$ (see also Altavilla et
al. 2004; Della Valle et al. 2005 for similar argument). Although
both SNe Ia in spiral and E/S0 galaxies show a wide range of $\Delta
m_{15}$, the fraction with $\Delta m_{15} < 1.2$ is 75$\%$ in
spirals and only 18$\%$ in E/S0 galaxies. The fact that faster
decliners (i.e., fainter SNe Ia) dominate in E/S0 galaxies is
apparently related to the age of the progenitors since E/S0 galaxies
are believed to have older low-mass stars. Recently, Sullivan et al.
(2006) demonstrated that the stretch factor of SNe Ia (another form
of $\Delta m_{15}$) in low- and high-redshift spiral galaxies bears
the same distribution at a confidence level of $95\%$, while the
distributions in elliptical galaxies are different as the faster
decliners (also the fainter SNe Ia) are usually seldom observed at
high redshift probably due to the selection effect and/or the SN
evolution effect. Since the decline rate $\Delta m_{15}$ is tightly
correlated with the stretch factor s, this difference found in
elliptical galaxies holds true for $\Delta m_{15}$.

Figure 2 shows the distribution of $\triangle m_{15}$ for different
types of SNe Ia. Compared with normal SNe Ia, SN 1991T-like events
have a flatter decline rate of 0.95$\pm$0.10 while SN 1991bg-like
objects have a much faster decline rate of 1.90$\pm$0.10. In the
figure SNe 2001ay and 2002cx are truly peculiar events. SN 2001ay
has the slowest decline rate in our sample, and Howell $\&$ Nugent
(2004) suggested that SN 2001ay is a SN Ia with the broadest light
curve. Li et al. (2003) noted that SN 2002cx has a pre-maximum
spectrum like SN 1991T (the classical hot SN Ia), a luminosity like
SN 1991bg (the classic subluminous event), a slow late-time decline,
and unidentified spectral lines (which is probably not a SN Ia as it
showed a spectroscopic feature extremely similar to those in some
faint type II-P SNe at nebular phase, private communication with Dr.
Weidong Li).

\begin{figure}
 \begin{minipage}[t]{0.5\textwidth}
    \centering
    \includegraphics[width=3in]{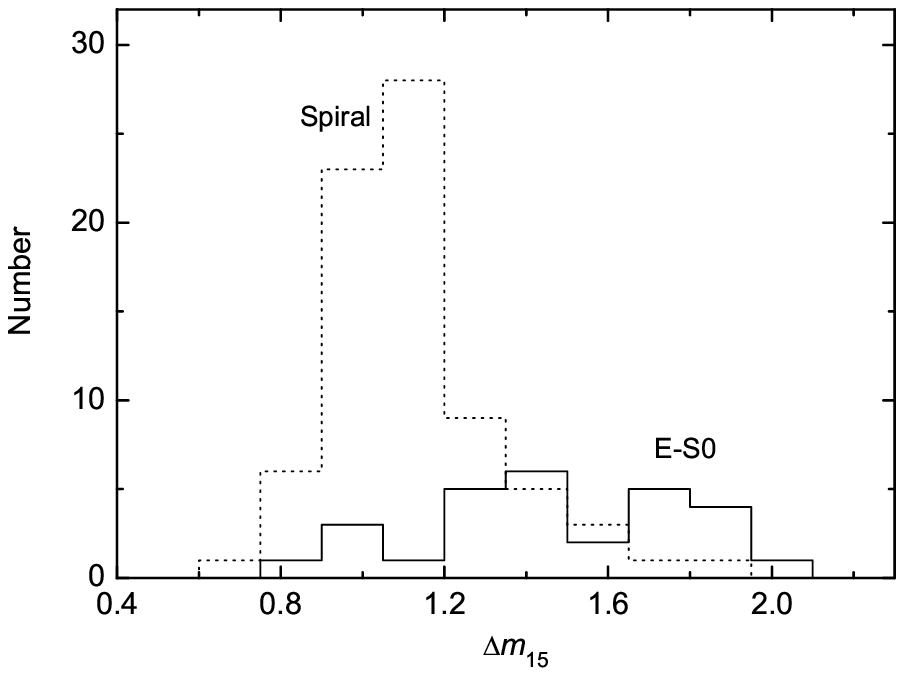}
    \vspace{-5mm}
    \caption{Distribution of $\triangle m_{15}$ for 28 SNe Ia in E/S0
   galaxies (solid line) and 77 in spiral galaxies (dotted line).}
    \label{Fig:fig1}
 \end{minipage} %
 \begin{minipage}[t]{0.5\textwidth}
   \centering
   \includegraphics[width=3in]{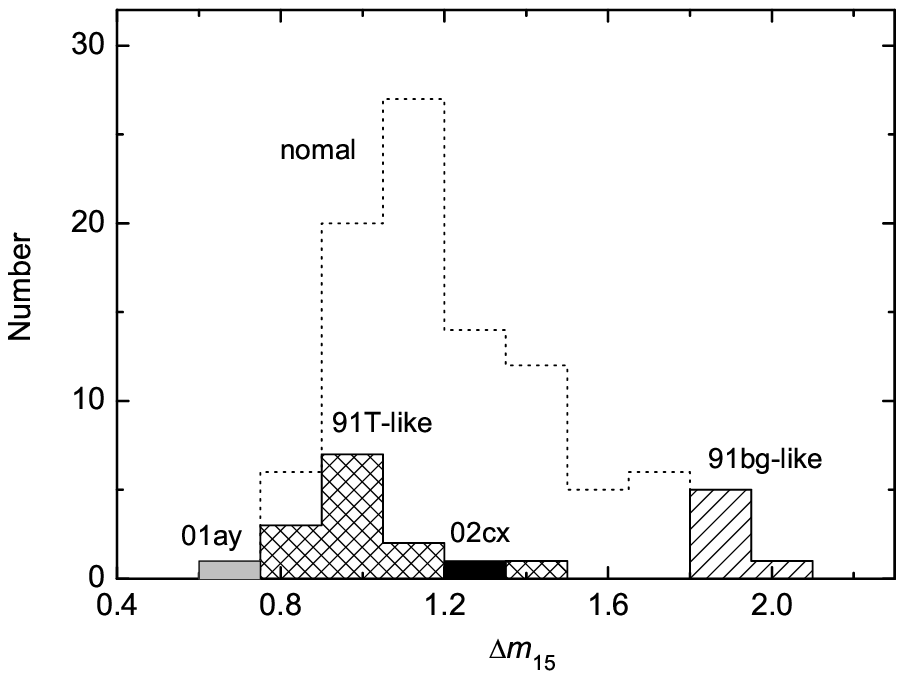}
   \vspace{-5mm}
   \caption{Distribution of $\triangle m_{15}$ for different SN Ia types. The
   dotted line shows the distribution of 90 normal SNe Ia. The cross-hatched
    area and the simply shaded area are for 13 SN 1991T-like events and 6 SN
    1991bg-like objects, respectively. Moreover, the gray-colored area and the
    black area show SNe 2001ay and 2002cx, respectively.}
   \label{Fig:fig2}
 \end{minipage} %
\end{figure}

\subsection{Distribution of $r_{\rm SN}/R_{25}$}
The distribution of SNe Ia in host galaxies can provide important
clues to their progenitors and give constraints on the explosion
models. In Figure 3, the decline rate $\triangle m_{15}$ is plotted
against the normalized radial distance $r_{\rm SN}/R_{25}$. The
filled circles mark the 26 SNe Ia in E/S0 galaxies, and the open
circles, the 70 SNe Ia in spirals, with the observational error bars
shown. In the figure SN 2002cs is a slow decliner (with $\triangle
m_{15}$=1.00) in the elliptical galaxy NGC 6702, and SN 2002dj is
another slow decliner (with $\triangle m_{15}$=1.05) in the
elliptical galaxy NGC 5018. Moreover, SN 1998es is a SN 1991T-like
object in a lenticular galaxy. Its host galaxy NGC 632 has a nuclear
starburst with a diameter of 18\,kpc, and the position of SN 1998es.

Figure 4 shows the distribution of the normalized radial distance
$r_{\rm SN}/R_{25}$. Wang (2002) first noticed that SNe Ia might be
more centrally concentrated in spiral galaxies than in E/S0
galaxies. In our study, the fraction of SNe Ia in E/S0 galaxies
located in the $r_{\rm SN}/R_{25}>1.0$ region is about 23$\%$, while
this fraction is only 4$\%$ for the SNe Ia in spirals. In other
words, SNe Ia in spiral galaxies could be more concentrated towards
the optical $R_{25}$ radius of their host galaxies than in E/S0
galaxies. A K-S test shows there is only a 33$\%$ probability that
the ancestors of SNe Ia in the E/S0 and the spiral galaxies have the
same radial distribution. Furthermore, the plot displays a
deficiency of SNe Ia in the central $r_{\rm SN}/R_{25}\leq 0.2$
region. The data on the SNe Ia radial distribution closer to center
the galaxies can be affected by absorption (Windhorst et al. 2002).
Note that this effect on the SNe Ia distribution closer to the
center of the galaxies may be more serious in spiral galaxies with
abundant gas and dust. Therefore, SNe Ia in spiral galaxies may be
more centrally located than the above number suggests.

\begin{figure}
 \begin{minipage}[t]{0.5\textwidth}
   \centering
    \includegraphics[width=3in]{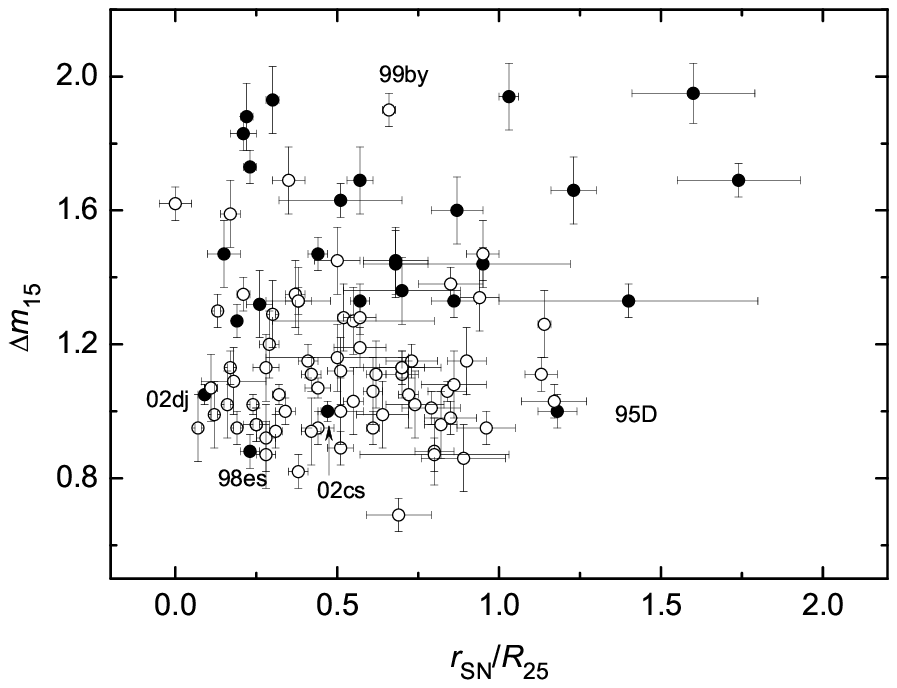}
    \vspace{-5mm}
    \caption{$\triangle m_{15}$ plotted against the relative radial
    distance $r_{\rm SN}/R_{25}$. The filled circles show 26 SNe Ia
    in E/S0 galaxies. The open circles represent 70 SNe Ia in spiral
    galaxies.}
    \label{Fig:fig3}
 \end{minipage} %
 \begin{minipage}[t]{0.5\textwidth}
   \centering
   \includegraphics[width=3in]{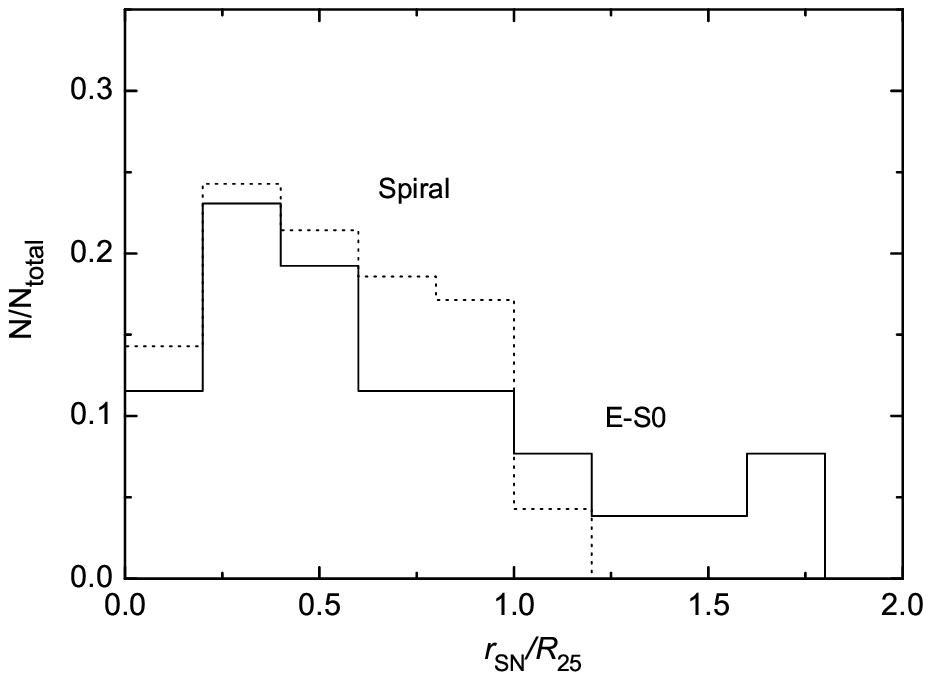}
   \vspace{-5mm}
   \caption{Distribution of the normalized radial distance $r_{\rm SN}/R_{25}$
   for SNe Ia in E/S0 galaxies (solid line) and in spiral galaxies (dotted line).}
   \label{Fig:fig4}
 \end{minipage} %
\end{figure}

\subsection{$M_{\rm Ni}$ -- $\triangle m_{15}$ Relation}
The mass of $^{56}$Ni produced during a SN Ia explosion is the
primary physical parameter determining the luminosity of the event.
One simple method to estimate the synthesized $^{56}$Ni mass is a
detailed calculation on the assumption that the maximum luminosity
of a SN Ia is proportional to the instantaneous rate of radioactive
decay (Arnett 1982; Arnett et al. 1985; Branch 1992). Thus the
implied bolometric luminosity at light maximum (the luminosity
integrated from ultraviolet to infrared) can be expressed as
\begin{equation}
  L_{\rm bol}=\alpha\dot{S}(t_{\rm R})M_{\rm Ni},
\end{equation}
where $M_{\rm Ni}$ is the total mass of $^{56}$Ni produced in the
explosion, and $\alpha$ is the ratio of bolometric to radioactivity
luminosity. $\dot{S}(r_{\rm R})$ is the radioactivity luminosity per
solar mass of $^{56}$Ni from its decay to $^{56}$Co and the
subsequent decay to $^{56}$Fe (e.g., Nadyozhin 1994):
\begin{equation}
  \dot{S}=6.45\times10^{43}e^{-t_{\rm R}/8.8}+1.45\times10^{43}e^{-t_{\rm
  R}/111.3}\,{\textrm{erg}\ \textrm{s}^{-1}\,M^{-1}_{\odot}},
\end{equation}
where $t_{\rm R}$ is the rise time in days from the SN explosion to
the light maximum. A typical rise time of $19\pm2$ days is assumed
for SNe Ia in our analysis (e.g., Branch 1992).

From the analytic solutions for the early-epoch SN Ia light curves,
Arnett (1982) found that $\alpha$ should be 1 exactly. On the other
hand, according to an early survey of SN Ia lightcurve calculations,
Branch (1992) found that $\alpha$ is slightly larger than unity. He
concluded $\alpha=1.2\pm0.2$ and noted that the value of $\alpha$ is
independent of the rise time. We assume $\alpha=1$ in our analysis
because the nickel produced above the photosphere may not make
significant contribution to the luminosity (e.g., Nugent et al.
1995). Stritzinger $\&$ Leibundgut (2005) noted that light curves of
two 3-D SNe Ia explosion models recently synthesized by the MPA
group are consistent with $\alpha=1$. Using Arnett's rule
($\alpha=1.0$) which states that the maximum luminosity of a SN Ia
is equal to the instantaneous energy deposition rate from the
radioactive decays within the expanding ejecta (Arnett 1982; Arnett
et al. 1985), and assuming a rise time to bolometric maximum of 19
days, one can derive from Equations (1) and (2) a simple formula for
the maximum luminosity of a SN Ia (e.g., Stritzinger $\&$ Leibundgut
2005):
\begin{equation}
  L_{\rm max}=(2.0\pm0.3)\times 10^{43}(\frac{M_{\rm Ni}}{M_{\odot}})\,\rm{erg\
  s^{-1}}.
\end{equation}
The uncertainty of the derived  $^{56}$Ni mass is primarily
determined by the errors in the adopted bolometric rise time and
the transformation efficiency of the trapped $\gamma$ energy.

It is generally believed that the $^{56}$Ni mass can be better
measured by the bolometric light curves constructed from the light
curves in the ultraviolet, optical and infrared wavelengths.
However, only a few SNe Ia have good observations covering from $UV$
to $IR$, which could be used as a training sample for $^{56}$Ni mass
determination of other SNe Ia. In Figure~5, the $^{56}$Ni masses
derived from 22 well-observed SNe Ia with $UBVRI$ and infrared light
curves, are plotted vs. $\triangle m_{15}$. For consistency in the
$\Delta m_{15}$ determinations, the $\triangle m_{15}$ values of the
22 training SNe Ia are taken from Table 1 in Wang et al. (2006). Of
the 22 SNe Ia with $^{56}$Ni estimates, 17 are from Table 2 in
Stritzinger et al. (2006b) and five (SNe 1995E, 1998aq, 1998de,
1999ac and 1999dq) from Table 1 in Stritzinger et al. (2006a). To
derive the $^{56}$Ni masses Stritzinger et al. (2006 a, b) made use
of  Equation (3). In Figure 5, the solid line corresponding to a
least squares fit of the data yields the following zero points and
dispersion:
\begin{equation}
  M_{\rm Ni}=(-0.61\pm0.09)\triangle m_{15}+(1.30\pm0.11)\ \ \ \sigma=0.13\ \ \ N=22,
\end{equation}
where the quoted dispersion ``$\sigma$'' is the simple root mean
square (rms) deviation of the points about the fit. Our fitting
result is very similar to that of  Mazzali et al. (2007). Note that
the least square fit line is similar to the Phillips
$M_{B}$--$\triangle m_{15}$ relation. A vertical deviation of
0.13\,M$_\odot$ (dotted lines) about the solid line is also shown.
It is suggested that the maximum luminosity of a SN Ia ($\triangle
m_{15}$) could be a direct consequence of the $^{56}$Ni mass via
modelling the light curves of SNe Ia (Arnett 1982; Cappellaro et al.
1997; Mazzali et al. 2001; Kasen 2006). The correlation between the
$^{56}$Ni mass and $\triangle m_{15}$ shown in Figure 5 may be taken
as observational evidence for this assumption.

\begin{figure}
   \begin{center}
   \mbox{\epsfxsize=0.7\textwidth\epsfysize=0.7\textwidth\epsfbox{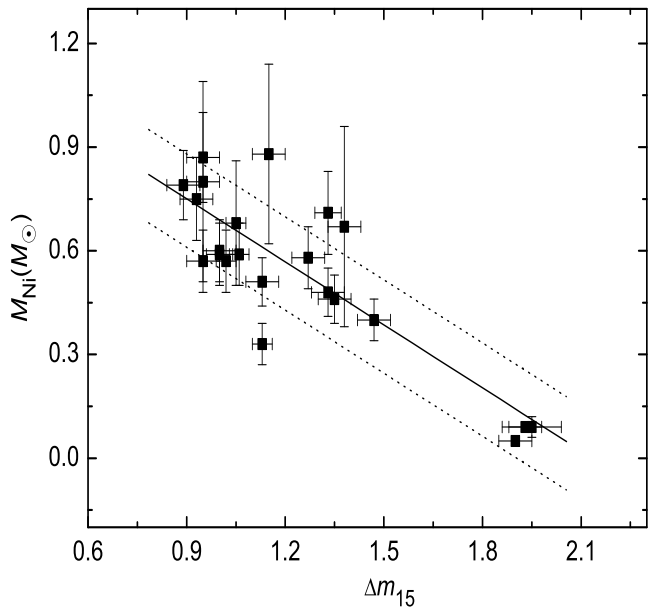}}
   \caption{$^{56}$Ni mass derived from maximum luminosity vs. $\triangle m_{15}$
    for 22 well-observed SNe Ia. The solid line marks the least squares fit. The
   dotted lines correspond to a vertical deviation of 0.13\,$M_\odot$ (i.e., a quoted
   dispersion ``$\sigma$'') about the solid line. Note that the solid line in the panel
   is similar to the relation obtained by Phillips.  }
   \label{Fig:fig5}
   \end{center}
\end{figure}

\subsection{Distribution of $^{56}$Ni Masses}
The amount of $^{56}$Ni synthesized from burning to nuclear
statistical equilibrium (NSE) in SN Ia explosions is probably the
most important physical parameter underlying the observed
correlation of SN Ia luminosities with their light curves. With
the $^{56}$Ni mass one can directly study the most sensitive part
of the explosion and provide constraints on the explosion
mechanisms (see the review Leibundgut 2000).  Here, according to
the empirical relation between the $^{56}$Ni mass and $\triangle
m_{15}$, i.e., Equation (4), we can estimate the $^{56}$Ni masses
and give the distribution of the $^{56}$Ni mass for a large number
of SNe Ia (see Figs. 6 and 7).

As shown in Figures 6 and 7, the $^{56}$Ni mass synthesized in SN Ia
explosions varies from 0.1 to 1.3\,$M_\odot$, with mean error
0.17\,$M_\odot$.  Figure~6 displays the distribution of the
$^{56}$Ni mass for SNe Ia in spiral and E/S0 galaxies, the black
area represents SN 2003fg. To produce SN 2003fg's unusual
luminosity, Howell et al. (2006) calculated, using Equation (1),
$^{56}$Ni mass, $M_{\rm Ni}=1.29\pm 0.07$\,$M_\odot$, a value that
was further confirmed by Jeffery et al. (2006). In Figure 7 we show
the distribution of the $^{56}$Ni mass for different SN Ia types,
and it is indicated that different SN Ia types (i.e., normal SNe Ia,
SN 1991bg-like events and SN 1991T-like objects) require the
existence of a range in the $^{56}$Ni masses in SN Ia explosions.

\begin{figure}
 \begin{minipage}[t]{0.5\textwidth}
   \centering
    \includegraphics[width=3in]{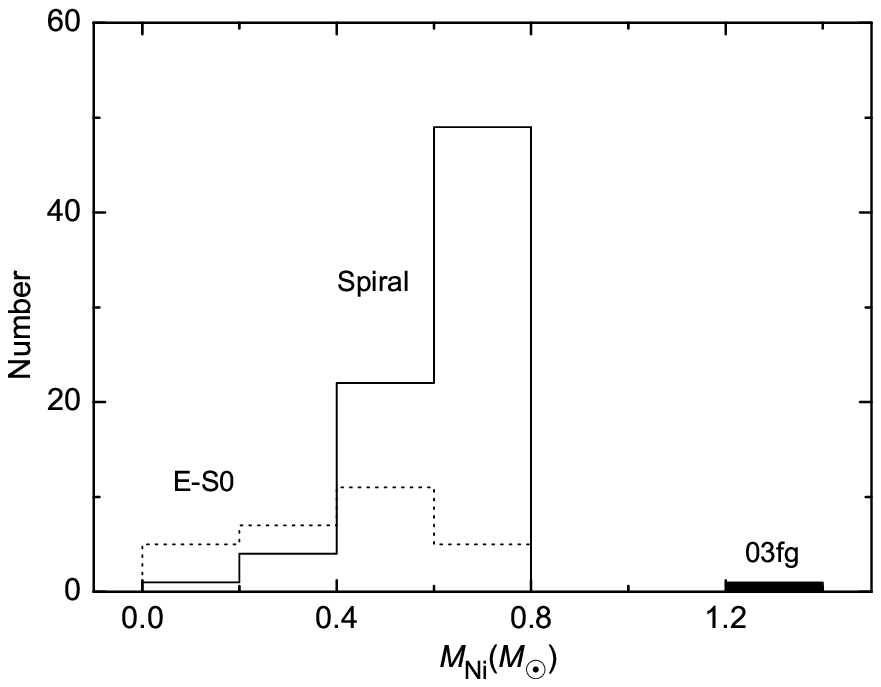}
    \vspace{-5mm}
    \caption{Distribution of the $^{56}$Ni mass for 28 SNe Ia in E/S0 galaxies
    (dotted line) and 76 in spiral galaxies (solid line) except for SN 2001ay
    which is truly peculiar. The black area represents SN 2003fg.}
    \label{Fig:fig6}
 \end{minipage} %
 \begin{minipage}[t]{0.5\textwidth}
   \centering
   \includegraphics[width=3in]{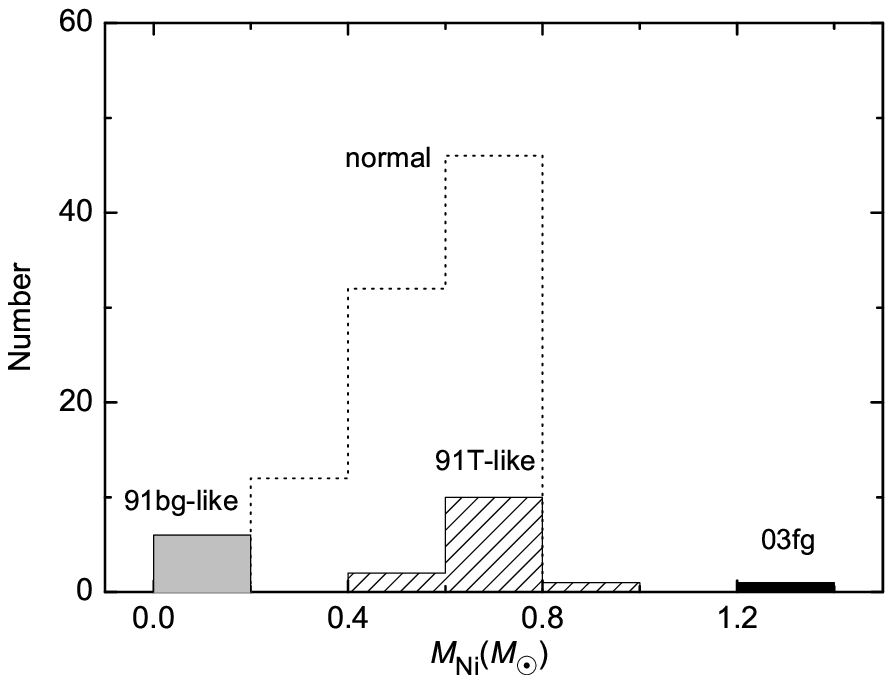}
   \vspace{-5mm}
   \caption{Distribution of the $^{56}$Ni mass for different SN Ia types. The dotted
   line shows the 90 normal SNe Ia. The gray-colored area and the single shaded area are
   for 6 SN 1991bg-like events and 13 SN 1991T-like objects, respectively. The black area
   represents SN 2003fg.}
   \label{Fig:fig7}
 \end{minipage} %
\end{figure}

\section{DISCUSSION AND CONCLUSIONS}
\label{sect:discussion}
As we know above (see Section 3.1), the brighter SNe Ia prefer to
occur in spiral galaxies and the fainter ones prefer to occur in
E/S0 galaxies. However, SNe 2002cs and 2002dj are slow decliners in
elliptical galaxies. This contradicts with previous conclusion that
elliptical galaxies produce only fainter SNe Ia. These two cases
probably suggest that the age may not be the exclusive factor
underlying SN Ia diversities. Moreover, SN 1998es is a slow decliner
in a nuclear starburst galaxy, which can probably be taken as an
instance of brighter events occurring in systems with ongoing star
formation (Sullivan et al. 2006). Figure 4 shows that SNe Ia might
be more concentrated towards the optical $R_{25}$ radius of their
host galaxies in spirals than in E/S0 galaxies. This suggests that
SNe Ia in spiral and E/S0 galaxies might have different progenitor
properties.

There is increasing evidence that a single parameter ($\triangle
m_{15}$) cannot account for the diversity in SN Ia (e.g., Benetti et
al. 2004). As shown in Figure 5, there is a relatively large scatter
$\sim$ 0.13\,$M_{\odot}$ in the $^{56}$Ni mass--$\triangle m_{15}$
relation. The dispersion may be partly due to uncertainties in the
distance and reddening adopted in the calculations, but an intrinsic
dispersion cannot be ruled out. Although it is believed that the
$^{56}$Ni mass in SN Ia explosions is the dominant physical
parameter affecting both the brightness and light curve shape (or
$\triangle m_{15}$), Woosley et al. (2006) proposed that some
parameters other than $M_{\rm Ni}$ also affect the light curves to
some extent, such as the total burned mass, the stable iron mass and
the degree of direct $^{56}$Ni mixing. However, since the empirical
$M_{\rm Ni}$--$\triangle m_{15}$ relation is similar to the Phillips
$M_{B}$--$\triangle m_{15}$ relation and the larger sample used in
the analysis helps to reduce the effect of random error, it is
reliable to use the relation to roughly estimate the $^{56}$Ni mass.

Based on the distribution of the $^{56}$Ni masses above, it is shown
that the $^{56}$Ni mass varies by at least a factor of ten (e.g.,
$M_{\rm Ni}=0.1-1.3$\,$M_{\odot}$). The explosions of SNe Ia might
produce a range in the amount of $^{56}$Ni synthesized from
$\sim$0.1\,$M_{\odot}$ associated with the subluminous objects to
$\sim$1.3\,$M_{\odot}$ for the most luminous events. The wide
distribution of the $^{56}$Ni masses raises the question as to what
physical mechanism(s) can account for this dispersion. During the
past years, a number of potential parameters in the
Chandrasekhar-mass deflagration scenario were investigated to answer
this question, e.g., C-O ratio, overall metallicity and central
density. However, 3D simulations by R\"{o}pke $\&$ Hillebrandt
(2004) indicated that different C-O ratios have a negligible effect
on the amount of $^{56}$Ni produced. Also R\"{o}pke et al. (2005b)
showed that there exits a $\sim20\%$ change in the resulting
$^{56}$Ni masses via altering the metallicity, which is consistent
with the analytic prediction of Timmes et al. (2003). Moreover, it
has been shown that changing the central density of the white dwarf
does influence the robustness of the explosion. Again, Stritzinger
et al. (2006a) suggested that it is unrealistic that any one of
these parameters, or even a combination of all three, can account
for a factor of ten range in the $^{56}$Ni mass.

Besides these parameters, the explosion mechanism itself is more
likely to influence the $^{56}$Ni masses synthesized in SN Ia
explosions (see Stritzinger $\&$ Leibundgut 2005). The standard
Chandrasekhar-mass model is the most popular explosion scenario for
SNe Ia. However, different Chandrasekhar-mass models can produce
different amount of $^{56}$Ni which is usually in the range of 0.4
-- 0.8\,$M_{\odot}$ (e.g., Mazzali et al. 2001), thus it is an
observational fact that a single class of Chandrasekhar-mass model
cannot account for the explosion properties of all the SNe Ia.
Recently, Mazzali et al. (2007) accomplished a systematic spectral
analysis of a large sample of well-observed SNe Ia and mapped the
velocity distribution of the main products of nuclear burning in SN
Ia explosions. They indicated that a single explosion scenario,
possibly a delayed detonation (DD) of Chandrasekhar-mass model, may
explain most SNe Ia. From the distribution of the $^{56}$Ni mass
displayed above, one can find that the $^{56}$Ni mass fall mostly in
the range of 0.4 -- 0.8\,$M_{\odot}$, thus we suggest that a
Chandrasekhar-mass model might account for normal SNe Ia which have
the $^{56}$Ni masses from 0.4\,$M_\odot$ to 0.8\,$M_\odot$.
Moreover, Sub-Chandrasekhar WDs are plausible candidates for the
subluminous objects. We note that one appealing advantage offered by
sub-Chandrasekhar model is the ability to obtain the progenitor
statistics predicted by population synthesis calculations (Livio
2000). In addition, the sub-Chandrasekhar model can easily explain
the observed diversity of SNe Ia by a one parameter sequence in
terms of the WD mass (Ruiz-Lapuente et al. 1995), and we also
suggested flexible progenitors for SN 1991bg-like events with
$M_{\rm Ni}$ $<$ 0.2\,$M_{\odot}$ (Cappellaro et al. 1997;
Stritzinger et al. 2006a, b). However, it must be noted that the
sub-Chandrasekhar model has difficulty in matching the observed
light curves and spectroscopy (e.g., H\"{o}flich $\&$ Khokhlov
1996).

As shown in Figure7, the $^{56}$Ni mass of SN 2003fg exceeds others
remarkably. Hillebrandt et al. (2007) argued that a lop-sided
explosion of a Chandrasekhar-mass WD could provide a better
explanation for SN 2003fg. They suggested that such objects must be
rare, since both a moderately high $^{56}$Ni mass
($\geq0.9$\,$M_\odot$) and rather special viewing direction are
rare, since both a moderately high $^{56}$Ni mass
($\geq0.9$\,$M_\odot$) and a rather special viewing direction are
rare. However, it is commonly believed that SN 2003fg could be from
a super-Chandrasekhar WD explosion (e.g., Uenishi et al. 2003; Yoon
$\&$ Langer 2005; Howell et al. 2006; Jeffery et al. 2006). It is
proposed that the influence of rotation of the accreting WDs may
cause the mass to exceed 1.4\,$M_\odot$. Mazzali et al. (2007) also
noted that some extremely luminous SNe Ia may come from very rapidly
rotating WDs whose mass exceeds Chandrasekhar mass, but these are
rare.

In summary, the wide distribution of the $^{56}$Ni mass suggests
that there are probably multiple origins of SN Ia explosion and/or
progenitor systems. We argue that sub-Chandrasekhar WDs could be the
progenitors of some extremely nickel-poor SNe Ia (with $M_{\rm Ni}$
$<$ 0.2\,$M_{\odot}$), and that super-Chandrasekhar WDs the
progenitors of some rare nickel-rich SNe Ia (with $M_{\rm Ni}$ $>$
0.8\,$M_{\odot}$). The Chandrasekhar-mass model remains the main
scenario for SN Ia explosions with a range in $^{56}$Ni masses from
0.4\,$M_\odot$ to 0.8\,$M_\odot$. To set further constraints on SN
Ia explosion/progenitor models, large samples of SNe Ia with
well-observed light curves and spectroscopy in nearby galaxies are
required to establish the connection of SN Ia properties with the
stellar environments of their host galaxies.

\begin{acknowledgements}
We acknowledge Dr. Weidong Li for SNe 2002cs and 2002dj parameters
before publications. We also thank the referee for careful reading
of the paper and valuable suggestions. This work is supported by the
National Natural Science Foundation of China (Grant Nos. 10433030,
10521001 and 10673007), the Foundation of the Chinese Academy of
Sciences (No. KJCX2-SW-T06), and Basic Research Funding at Tsinghua
University (JCqn2005036).
\end{acknowledgements}

\label{lastpage}
\end{document}